\documentclass[aps,amsmath,amssymb,prl,twocolumn,10pt,showpacs,superscriptaddress,letterpaper]{revtex4-1}
\usepackage{graphicx}
\usepackage{bm}
\usepackage{txfonts}
\usepackage{mathrsfs}
\usepackage{feynmf}
\usepackage{comment}
\usepackage{epsfig}
\usepackage{setspace}
\usepackage{mathtools}

\newcommand{\al}{\alpha}

\newcommand{\de}{\delta}

\newcommand{\ka}{\kappa}
\newcommand{\la}{\lambda}

\newcommand{\x}{\chi}

\newcommand{\W}{\Omega}

\newcommand{\pd}{\partial}

\newcommand{\round}[1]{\left( #1 \right)}
\renewcommand{\square}[1]{\left[ #1 \right]}
\newcommand{\curly}[1]{\left\{#1\right\}}

\newcommand{\ang}[1]{\left\langle #1 \right\rangle}

\newcommand{\beq}{\begin{equation}}
\newcommand{\eeq}{\end{equation}}
\newcommand{\Beq}{\begin{eqnarray}}
\newcommand{\Eeq}{\end{eqnarray}}
\newcommand{\bml}{\begin{multline}}

\newcommand{\bea}{\begin{align}}
\newcommand{\ena}{\end{align}}
\newcommand{\bsp}{\begin{split}}
\newcommand{\esp}{\end{split}}
\newcommand{\down}{\downarrow}
\newcommand{\up}{\uparrow}

\newcommand{\tal}{\tilde\alpha}

\newcommand{\bS}{{\boldsymbol{S}}}

\newcommand{\ex}{{\boldsymbol e_x}}
\newcommand{\ey}{{\boldsymbol e_y}}
\newcommand{\ez}{{\boldsymbol e_z}}

\newcommand{\bj}{{\boldsymbol j}}
\newcommand{\bJ}{{\boldsymbol J}}

\renewcommand{\bm}{{\boldsymbol m}}

\newcommand{\bn}{{\boldsymbol n}}

\newcommand{\sF}{\mathscr{F}}
\newcommand{\sA}{\mathscr{A}}
\newcommand{\sV}{\mathscr{V}}

\newcommand{\ve}{\varepsilon}

\newcommand{\bb}{\boldsymbol{b}}

\begin{document}
\title{Antiferromagnet-Mediated Spin Transfer Between Metal and Ferromagnet}
\author{So Takei}
\affiliation{Department of Physics and Astronomy, University of California, Los Angeles, California 90095, USA}
\author{Takahiro Moriyama}
\affiliation{Institute for Chemical Research, Kyoto University, 611-0011 Uji, Kyoto, Japan}
\author{Teruo Ono}
\affiliation{Institute for Chemical Research, Kyoto University, 611-0011 Uji, Kyoto, Japan}
\author{Yaroslav Tserkovnyak}
\affiliation{Department of Physics and Astronomy, University of California, Los Angeles, California 90095, USA}
\date{\today}

\begin{abstract}
We develop a theory for spin transported by coherent N\'eel dynamics through an antiferromagnetic insulator coupled to a ferromagnetic insulator on one side and a current-carrying normal metal with strong spin-orbit coupling on the other. The ferromagnet is considered within the mono-domain limit and we assume its coupling to the local antiferromagnet N\'eel order at the ferromagnet$|$antiferromagnet interface through exchange coupling. Coupling between the charge current and the local N\'eel order at the other interface is described using spin Hall phenomenology. Spin transport through the antiferromagnet, assumed to possess an easy-axis magnetic anisotropy, is solved within the adiabatic approximation and the effect of spin current flowing into the ferromagnet on its resonance linewidth is evaluated. Onsager reciprocity is used to evaluate the inverse spin Hall voltage generated across the metal by a dynamic ferromagnet as a function the antiferromagnet thickness.
\end{abstract}
\maketitle

\singlespacing

Spintronics of antiferromagnets (AFs), where AFs take on the role of the central active component, is identified as one of the most important emerging topics in the field of magnetism today~\cite{macdonalPTRSA11,*duineNATM11,*brataasNATM12}. Robustness to magnetic perturbations due to their total magnetic compensation, as well as characteristic dynamical scale in the THz range may render AFs advantageous over ferromagnets (Fs) for spintronics device applications. In addition, recent works on AFs have shown that the important phenomena responsible for the success of F-based spintronics also have AF counterparts, giving added impetus for AF-based spintronics research. Indeed, giant magnetoresistance and current-induced torques~\cite{nunezPRB06,*haneyPRB07,*urazhdinPRL07,*weiPRL07,*haneyPRL08,*xuPRL08,*gomonayPRB10,*gomonayPRB12}, anisotropic magnetoresistance~\cite{shickPRB10,*parkNATM11} and spin superfluidity~\cite{konigPRL01,*soninAP10,*takeiPRB14}, as well as current-induced domain wall motion~\cite{wieserPRL11,*kimPRB14} and coupled dynamics between conduction electrons and background magnetic texture~\cite{halsPRL11,*swavingPRB11,*chengPRB12,*tvetenPRL13}, are all shown to be possible in AFs as well.

An important aspect of AF-based spintronics is the use of AFs as a medium to transport spin angular momentum. Spin transfer through AFs has been the focus of several recent experimental endeavors. Both Hahn {\em et al}.~\cite{hahnEPL14} and Wang {\em et al}.~\cite{wangPRL14} demonstrated spin transport through an AF insulator, NiO, using an YIG$|$NiO$|$Pt heterostructure (YIG standing for the insulating ferrimagnet yttrium iron garnet). Inverse spin Hall signal showed robust spin pumping from YIG into Pt even in the presence of the intervening NiO, suggesting efficient spin transport through the AF. More recently, Moriyama {\em et al}. used spin-torque ferromagnetic resonance (ST-FMR) to demonstrate the propagation of spin excitations through a metallic AF, IrMn, using a Pt$|$IrMn$|$CoFeB trilayer~\cite{moriyamaCM14} as well as NiO using a Pt$|$NiO$|$FeNi trilayer~\cite{moriyamaCM15}. Spin current injected from the Pt was shown to change the FMR linewidth, also suggesting the transfer of spin angular momentum through the central AF. Given the rising interest in AF spintronics and the recent experimental focus, a theoretical account of spin transport through an experimentally relevant normal metal (N)$|$AF$|$F trilayer is highly desirable.

\begin{figure}[t]
\centering
\includegraphics*[width=0.9\linewidth]{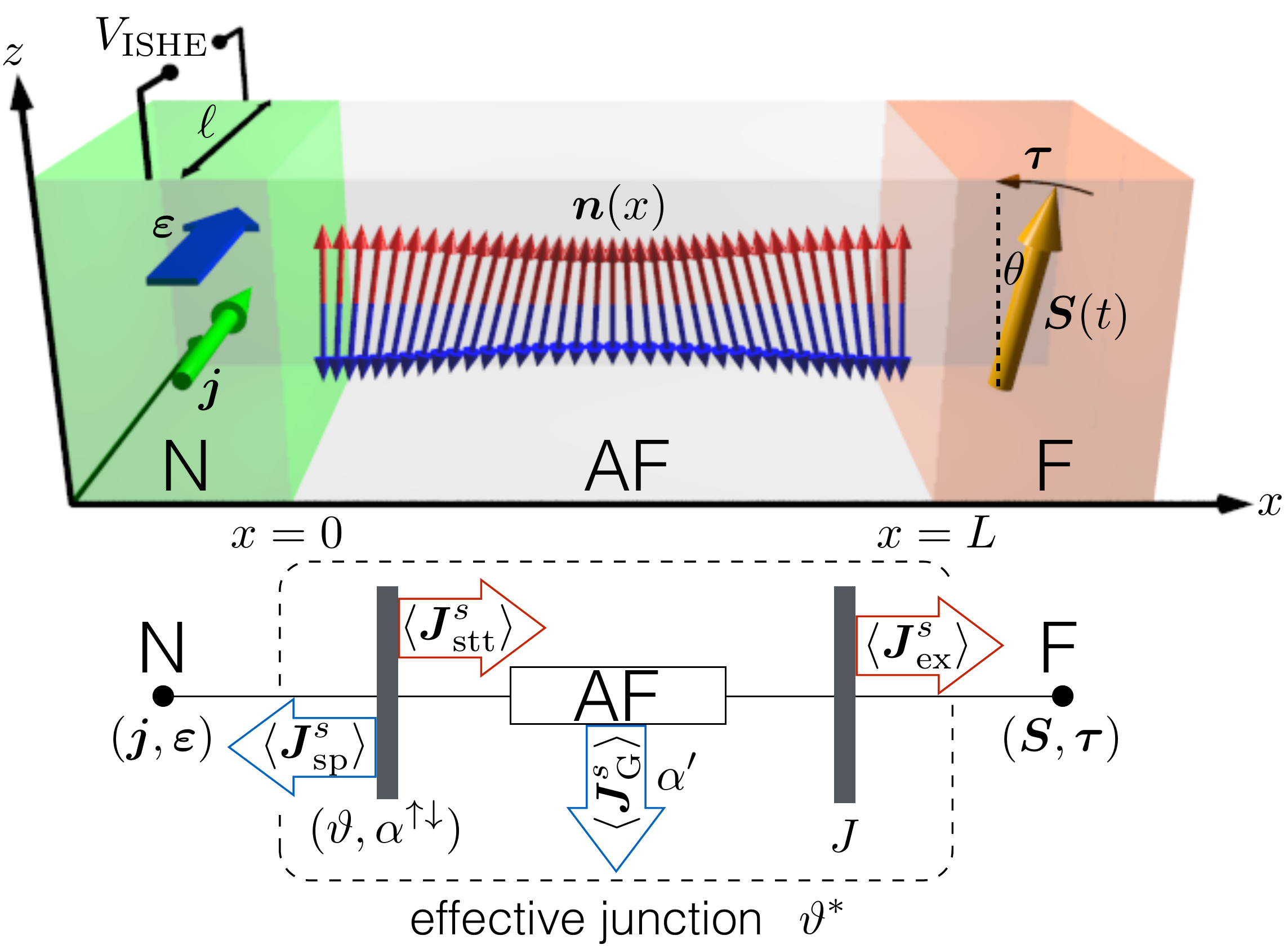}
\caption{Normal-metal (N)$|$antiferromagnet (AF)$|$ferromagnet (F) trilayer considered in this work. N sustains a dc charge current $\bj$ and F is described by a time-dependent macrospin $\bS(t)$. Spin transfer $\langle\bJ^s_{\rm ex}\rangle$ occurs via the exchange coupling $J$ at the AF$|$F interface, while spin transfer across the AF$|$N interface has a spin transfer torque contribution $\langle\bJ^s_{\rm stt}\rangle$ (proportional to the effective interfacial spin Hall angle $\vartheta$) and a spin pumping contribution $\langle\bJ^s_{\rm sp}\rangle$ (proportional to the interfacial spin-mixing conductance $\al^{\up\down}$). The AF Gilbert damping, parametrized by $\al'$, leads to the loss of spin current $\langle\bJ^s_{\rm G}\rangle$ in the AF bulk. The central AF can be thought of as an effective interface that couples $\bj$ and $\bS$ with an effective spin Hall angle $\vartheta^*$.}
\label{setup}
\end{figure}

In this Letter, we develop a general phenomenology for spin transport through an AF by collective N\'eel order parameter dynamics, focusing on an N$|$AF$|$F trilayer relevant for both the spin-pumping/inverse spin Hall as well as the ST-FMR experiments mentioned above (see Fig.~\ref{setup}). Spin Hall phenomenology, applicable to a wide range of different AF$|$N interfaces obeying certain structural/crystalline symmetries, is utilized to model the spin transfer at the AF$|$N interface, while the exchange coupling is assumed at the AF$|$F interface. As one of the main achievements of this work we develop a simple ``circuit" model, a pictorial visualization of spin flow, that allows one to keep track of spin transfer through various parts of the heterostructure (see bottom half of Fig.~\ref{setup}). From the circuit model, we see that spin is both injected into (i.e., $\langle\bJ^s_{\rm stt}\rangle$) and ejected (i.e., $\langle\bJ^s_{\rm sp}\rangle$) out of the AF at the AF$|$N interface due to spin Hall/spin-torque effects and spin pumping, respectively. The collective N\'eel dynamics leads to Gilbert damping and to the loss of spin current (i.e., $\langle\bJ^s_{\rm G}\rangle$) within the AF bulk, and the exchange coupling at the AF$|$F interface leads to spin transfer (i.e., $\langle\bJ^s_{\rm ex}\rangle$) across the interface. We first study how spin transport through the AF modifies the linewidth at FMR, akin to the ST-FMR~\cite{moriyamaCM14,moriyamaCM15}. The FMR linewidth is quantified in terms of the effective spin Hall angle and spin-mixing conductance at the AF$|$N interface, the exchange coupling at the AF$|$F interface, as well as AF Gilbert damping. We show that linewidths, measured for different electrical currents in N and AF thicknesses, can be used to extract the effective spin Hall angle and spin-mixing conductance at the N$|$AF interface, as well as the bulk Gilbert damping. By invoking Onsager reciprocity, we also make connections with the inverse spin Hall experiments~\cite{hahnEPL14,wangPRL14} and compute the inverse spin Hall voltage generated across N arising as a result of a dynamic F macrospin.

As shown in Fig.~\ref{setup}, an insulating AF is attached on one side to a mono-domain F and on the other to a paramagnetic N with strong spin-orbit coupling. The N and F sustain dc charge current density $\bj$ and a time-dependent macrospin $\bS(t)$, respectively. We consider a bipartite AF, which can be characterized by two hydrodynamic variables, $\bn(x,t)$ and $\bm(x,t)$, parametrizing the staggered (N\'eel) and smooth (magnetic) components of the spins, respectively~\cite{baryakhtarSJLTP79,*haldanePRL83,*auerbachBOOK94}. We assume easy-axis magnetic anisotropy along the $z$ axis in the AF, as well as full translational and rotational symmetries in the $yz$ planes so that our treatment essentially reduces to a one-dimensional problem that depends only on the coordinate $x$. The free energy $\sF$ for the AF and its coupling to the F reads
\beq
\begin{aligned}
\label{f}
\sF&=\int_0^L dx\curly{\frac{A}{2}[\pd_x\bn(x)]^2+\frac{\bm(x)^2}{2\x}-\frac{\ka}{2}n_z(x)^2}\\
&\qquad\qquad\qquad\qquad\qquad\qquad\qquad-J\bS\cdot\bn(L)\, ,
\end{aligned}
\eeq
where $A$ and $\x$ are the N\'eel order stiffness and spin susceptibility, respectively, $\ka$ the uniaxial anisotropy parameter, and $J$ the exchange coupling between AF and F. 

The Landau-Lifshitz-Gilbert dynamics in the bulk AF corresponding to Eq.~\eqref{f} can be written as
\begin{align}
\label{ndotG}
s(\dot\bn+\al\bn\times\dot\bm)&=\chi^{-1}\bm\times\bn\, ,\\
\label{mdotG}
s(\dot\bm+\al\bm\times\dot{\bm}+\al'\bn\times\dot\bn)&=\bn\times(A\pd_x^2\bn+\ka n_z\ez)\,,
\end{align}
where $\al$ and $\al'$ are (independent) Gilbert damping parameters and $s$ is the roughly saturated spin density (per unit length)~\footnote{The saturated spin density is defined by $s=\hbar S_0\sA/\sV$, where $S_0$ is the magnitude of each AF spin, $\sA$ is the cross-sectional area (along the $yz$ plane) of the AF, and $\sV$ is the volume occupied by each spin.}. In the limit of slow dynamics (i.e., $\tau\gg \hbar/E_{\rm ex}$, where $\tau$ is the time scale for AF dynamics and $E_{\rm ex}$ is the AF exchange energy scale) and strong local N\'eel order (i.e., $|\bm(x,t)|\ll 1$), one may first use Eq.~\eqref{ndotG} to solve for $\bm$, insert the solution into Eq.~\eqref{mdotG}, and arrive at the following dynamics for the AF N\'eel vector, 
\beq
\label{ndyn}
\x s^2\bn\times\ddot\bn+s\al'\bn\times\dot\bn\approx A\bn\times\pd_x^2\bn+\ka\bn\times n_z\ez\, .
\eeq

The AF spins are excited by the current and the dynamic macrospin. The effects of these external perturbations are localized at the interfaces and thus enter the AF dynamics as boundary conditions. At the AF$|$N interface, spin currents arise via spin transfer torque and spin-pumping, both of which can be accounted for using spin Hall phenomenology~\cite{tserkovnyakPRB14}. Based on structural symmetries at the interface, spin Hall phenomenology allows us to write down a general expression for the spin transfer torque that applies to a variety of F- and AF-based heterostructures with different microscopic details. In the presence of full translational and rotational symmetries in the $yz$ plane and with the breaking of reflection symmetry along the $x$ axis, there are two contributions to the spin current (integrated over the interface area) flowing into AF~\cite{tserkovnyakPRB14}
\beq
\label{jsl}
\bJ^s_l=[\vartheta\bn\times(\ex\times\bj)\times\bn-\hbar\al^{\up\down}\bn\times\dot\bn]|_{x=0}\equiv\bJ^s_{\rm stt} -\bJ^s_{\rm sp}\, ,
\eeq
where the first term is the so-called spin Hall-like (dissipative) contribution and the second term describes spin-pumping. The coefficient $\vartheta$ is proportional to the (tangent of the) effective spin Hall angle at the AF$|$N interface~\cite{tserkovnyakPRB14}; although $\vartheta$ can, in general, depend on the orientation of $\bn$, we will treat it as a constant here. We will also disregard any anisotropies of $\al^{\up\down}$ with respect to the orientations of $\bn$ and $\dot\bn$, assuming that the exchange energy scale at the interface dominates over the energy scale of spin-orbit interactions. 

While the coefficient $\al^{\up\down}$ is proportional to the real part of the (generally complex) spin-mixing conductance $g^{\up\down}$ for the AF$|$N interface~\cite{tserkovnyakPRB14}, its imaginary part gives rise to a term $\propto\dot\bn$ in Eq.~\eqref{jsl}. The assumed structural symmetry, in principle, also allows for the so-called field-like (reactive) contribution given by $\eta(\ex\times\bj)\times\bn$.  Assuming the magnitudes of all four terms are small (as compared to $\sqrt{A\kappa}$ to be precise) we will compute the FMR linewidth to linear-order in the coefficients $\vartheta$, $\eta$ and $g^{\up\down}$. Since the terms that do not appear in Eq.~\eqref{jsl} are non-dissipative terms, they do not enter the linewidth at linear-order in these coefficients. We will, therefore, drop these contributions from the following discussion and only consider the two terms in Eq.~\eqref{jsl}.

The exchange coupling $J$ [given by the last term in Eq.\eqref{f}] leads to the following expression for spin current flowing across the AF$|$F interface,
\beq
\label{jsr}
\bJ^s_r\equiv \bJ^s_{\rm ex}=\frac{\pd\sF}{\pd\bS}\times\bS=J\bS\times\bn|_{x=L}\, .
\eeq
The spin current in the AF bulk can be read off from Eq.~\eqref{mdotG} (dropping the Gilbert damping and anisotropy terms) and the resulting continuity equation [i.e., $s\dot\bm=-\pd_x(-A\bn\times\pd_x\bn)$], so that $\bJ^s_{\rm AF}(x)=-A\bn(x)\times\pd_x\bn(x)$. The continuity of spin current across each interface then leads to the boundary conditions
\beq
\label{bcs}
\bJ^s_l=\bJ^s_{\rm AF}(x=0)\, ,\quad\bJ^s_r=\bJ^s_{\rm AF}(x=L)\, .
\eeq

\begin{figure}[t]
\centering
\includegraphics*[width=0.9\linewidth]{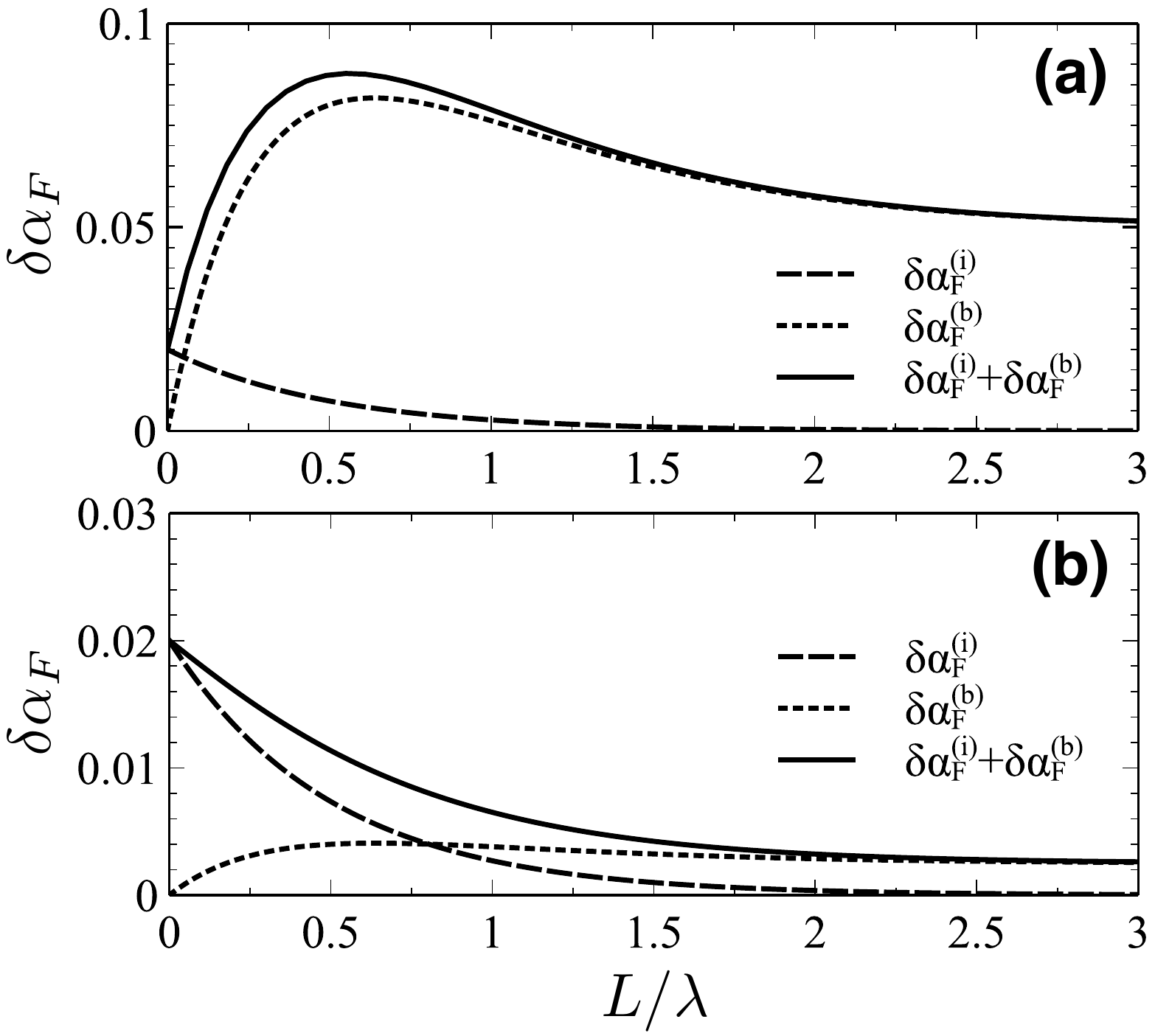}
\caption{The interfacial contribution $\de\al_F^{(i)}$ (dashed lines), the bulk contribution $\de\al_F^{(b)}$ (dotted lines) and the total contribution $\de\al_F$ (solid lines) to the extrinsic FMR linewidth ($S$ set to unity) are plotted as a function of the (normalized) system size $L/\la$. We fix the following parameters: $\eta=1$, $\vartheta j_y/b_0=0.01$ and $\al^{\up\down}=0.01$. Two regimes are considered for the AF Gilbert damping $\tal$: (a) the strong damping regime $\tal=0.2$; and (b) the weak damping regime $\tal=0.01$ (see text for more details).}
\label{plot}
\end{figure}

The dynamic N\'eel texture $\bn(x,t)$ can be obtained using the low-frequency (adiabatic) approximation, valid in the regime $\W\ll\W_0$, where $\W$ and $\W_0$ are the FMR and the AF resonance frequencies, respectively. Within this approximation, the AF N\'eel texture is first solved for an arbitrary static $\bS$, the result denoted by $\bn^{(0)}(x,\bS)$. Since $\bS(t)$ varies sufficiently slowly in time compared to the characteristic AF time scale, the N\'eel texture in the adiabatic limit will arrange itself into the static configuration corresponding to $\bS(t)$ at every moment in time and is given by $\bn(x,t)\approx\bn^{(0)}[x,\bS(t)]\equiv\bn^{(0)}(x,t)$. The above calculation does not account for spin current losses due to the AF dynamics (i.e., spin-pumping at the AF$|$N interface and Gilbert damping in the AF bulk). Taking these losses into account up to linear-order corrections to the adiabatic result, the spin current $\langle\bJ^s_{\rm ex}\rangle$ entering F, time-averaged over a cycle of FMR precession (the angle brackets $\ang{\cdots}$ hereafter representing time-average over a cycle of FMR precession), is given by $\langle\bJ^s_{\rm ex}\rangle=\langle\bJ^s_{\rm stt}\rangle-\langle\bJ^s_{\rm sp}\rangle-\langle\bJ^s_{\rm G}\rangle$ (c.f. Fig.~\ref{setup}), where the spin-transfer torque contribution is given by inserting the adiabatic result for the N\'eel texture into Eq.~\eqref{jsr}
\beq
\label{jssh}
\langle\bJ^s_{\rm stt}\rangle=J\langle\bS(t)\times\bn^{(0)}(L,t)\rangle\, ,
\eeq
and the loss terms read
\beq
\begin{aligned}
\label{jsdamp}
\langle\bJ^s_{\rm sp}\rangle&=\hbar\al^{\up\down}\langle\bn^{(0)}(0,t)\times\dot\bn^{(0)}(0,t)\rangle\, ,\\
\langle\bJ^s_{\rm G}\rangle&=s\al'\int_0^Ldx\  \langle\bn^{(0)}(x,t)\times\dot\bn^{(0)}(x,t)\rangle\, .
\end{aligned}
\eeq
The first term in Eq.~\eqref{jsdamp} describes (time-averaged) spin current lost due to spin pumping at the AF$|$N interface and the second term corresponds to Gilbert damping in the AF bulk. 

An analytical result for the FMR linewidth can be obtained if we consider small deviations of $\bS(t)$ away from the $z$ axis (parallel to the static FMR field and the AF easy-axis); we take $\bj=j_y\ey+j_z\ez$ and assume $|\bj|$ to be weak such that a linear-response treatment is sufficient. In this case, the N\'eel unit vector $\bn$ should not deviate far from the $z$ axis and we may evaluate the above results with respect to small transverse fluctuations, i.e., $\bS(t)\approx S[s_x(t),s_y(t),1]$ and $\bn(x,t)\approx [n_x(x,t),n_y(x,t),1]$ with $|s_x(t)|,|s_y(t)|\ll1$ and $|n_x(x,t)|,|n_y(x,t)|\ll1$. Within this treatment, the transverse components $\bn_\perp=(n_x,n_y)^T$ obey $\x s^2\ddot \bn_\perp+s\al\dot\bn_\perp=A\pd_{x}^2 \bn_\perp-\ka \bn_\perp$ [c.f. Eq.~\eqref{ndyn}], and $\bn^{(0)}(x,t)$ has the form 
\beq
\label{n0}
\bn^{(0)}\approx\ez+f(x)[\ez\times\bS(t)]\times\ez+g(x)\ez\times\bS(t)+h(x)\ex\, ,
\eeq
where the functions $f(x)$ and $g(x)$ (to linear-order in the current) are given by
\begin{align}
f(x)&=\frac{1}{S}\frac{\cosh\frac{x}{\la}}{\cosh\frac{L}{\la}+\frac{1}{\eta}\sinh\frac{L}{\la}}\, , \\
g(x)&=\frac{1}{S}\frac{\sinh\frac{L-x}{\la}+\frac{1}{\eta}\cosh\frac{L-x}{\la}}{\round{\cosh\frac{L}{\la}+\frac{1}{\eta}\sinh\frac{L}{\la}}^2}\frac{\vartheta j_y}{\sqrt{A\kappa}}\, ,
\end{align}
and $h(x)\propto\vartheta j_z$ is not explicitly shown here since this term will not contribute to the linewidth within the current theoretical treatment. Here, $\eta\equiv JS/\sqrt{A\ka}$, and $\la\equiv\sqrt{A/\ka}$ is the AF healing length. 

The spin current $\bJ^s_{\rm ex}$ entering F modifies the F dynamics as
\beq
\label{sdyn2}
\hbar\dot\bS=\bb\times\bS-\frac{\hbar\al_F}{S}\bS\times\dot\bS+\bJ^s_{\rm ex}\, ,
\eeq
where $\al_F$ is the intrinsic Gilbert damping parameter in F and $\bb=-b_0\ez$ is the static FMR field (in units of energy). Inserting Eq.~\eqref{n0} into Eqs.~\eqref{jssh} and \eqref{jsdamp} and performing the time-average over the last two terms in Eq.~\eqref{sdyn2}, the full FMR linewidth can be read off directly by summing the coefficients appearing in front of $\langle\bS\times\dot\bS\rangle$. The total Gilbert damping parameter is then given by $\al'_F=\al_F+\de\al_F^{(i)}+\de\al_F^{(b)}\equiv\al_F+\de\al_F$, where the extrinsic contribution $\de\al_F$ has the interfacial contribution $\de\al_F^{(i)}$ and the AF bulk contribution $\de\al_F^{(b)}$: 
\begin{align}
\label{dali}
\de\al_F^{(i)}&=\frac{1}{S}\frac{\round{\frac{\vartheta j_y}{b_0}+\al^{\up\down}}}{\round{\cosh\frac{L}{\la}+\frac{1}{\eta}\sinh\frac{L}{\la}}^2}\, ,\\
\label{dalb}
\de\al_F^{(b)}&=\frac{\tal}{S}\frac{\frac{L}{\la}+\frac{1}{2}\sinh\frac{2L}{\la}}{\round{\cosh\frac{L}{\la}+\frac{1}{\eta}\sinh\frac{L}{\la}}^2}\, ,
\end{align}
where $\tal=s\al'\la/2\hbar$. The former originates from spin injection and spin-pumping at the AF$|$N interface while the latter from Gilbert damping in the AF bulk. Eqs.~\eqref{dali} and \eqref{dalb} constitute the main result of this work.

As seen from Eqs.~\eqref{dali} and \eqref{dalb}, the healing length $\la$ sets the distance over which spin propagation decays inside the AF. The healing length is determined from the slope of the linewidth vs. $j_y$ curves for various thicknesses $L$ and by extracting the decay length. It is important to note that the current theory only considers spin transport mediated by coherent N\'eel dynamics, and does not take account of spin transported by incoherent thermal magnons. The latter contribution is suppressed at sufficiently low temperatures by some power of the ratio $T/T_N$, where $T_N$ is the N\'eel ordering temperature of the AF, and since magnon-mediated transport is expected to decay over the spin diffusion length $\la_{\rm sd}$, it is strongly suppressed for $\la_{\rm sd}\ll L$. 

Once the AF healing length is known, the parameters $\vartheta$, $\al^{\up\down}$ and $\tal$ can be extracted by measuring the FMR linewidth for various $j_y$ and $L$. While the effective spin Hall angle $\vartheta$ can be obtained from the slope of a linewidth vs. $j_y$ curve, the Gilbert damping parameter $\tal$ can be extracted in the regime $L\gg\la$, in which the linewidth depends only on $\tal$ (see Fig.~\ref{plot}),
\beq
\label{dalF8}
\de\al_F\overset{\frac{L}{\la}\rightarrow\infty}\rightarrow\frac{\tal}{S}\round{\frac{\eta}{1+\eta}}^2\equiv\de\al_F^\infty\, .
\eeq
For $L\ll\la$, we expand $\de\al_F$ to linear order in $L/\la$,
\beq
\begin{aligned}
\label{c0c1}
\de\al_F&\approx\frac{1}{S}\round{\frac{\vartheta j_y}{b_0}+\al^{\up\down}}+\frac{1}{S}\square{\tal-\frac{2}{\eta}\round{\frac{\vartheta j_y}{b_0}+\al^{\up\down}}}\frac{L}{\la}\\
&\equiv c_0+c_1\frac{L}{\la}\, ,
\end{aligned}
\eeq
from which we see that $\al^{\up\down}$ can be extracted at zero current (i.e., $j_y=0$) and measuring the linewidth for $L\ll\la$.

The extrinsic linewidth exhibits qualitatively different behavior depending on the relative magnitudes of the bulk and the interfacial contributions (see Fig.~\ref{plot}). For strong Gilbert damping ($c_1>0$) the bulk damping in the AF dominates over the interface effects and the linewidth grows initially as $L$ increases, saturating eventually as $L/\la\rightarrow\infty$ [see Fig.~\ref{plot}(a)]. In the limit of weak Gilbert damping [see Fig.~\ref{plot}(b)], i.e., $c_1<0$ (and $\de\al_F^\infty<c_0$), $\de\al_F$ exponentially decays as $L$ increases. In Ref.~\onlinecite{wangPRL14}, the FMR linewidth was measured for various AF thicknesses in the absence of the electrical current. The gradual increase in the linewidth obtained there as a function of the thickness is more consistent with our strong Gilbert damping regime [c.f. Fig.~\ref{plot}(a)]. 

Moriyama {\em et al}.~\cite{moriyamaCM15} has recently reported an ST-FMR measurement for single crystal NiO using a Pt$|$NiO$|$FeNi trilayer, observed a linear dependence of the FMR linewidth on the Pt electrical current, and extracted the slope [c.f. Eq.~\eqref{dali}]. The reported ratio $\beta\approx0.82$ between the slopes for the Pt$|$NiO$|$FeNi and Pt$|$FeNi samples indicated relatively efficient spin transfer through NiO compared to previous reports~\cite{hahnEPL14,wangPRL14}, and was attributed to the single crystal nature of the AF layer. Applying Eq.~\eqref{dali} directly to the experiment (see also Ref.~\onlinecite{moriyamaCM15}), the theoretical ratio $\beta_{\rm th}$ (ignoring a possible difference in $\vartheta$ between the two samples) is given by $\beta_{\rm th}=[\cosh(L/\la)+\sinh(L/\la)/\eta]^{-2}$. Using $\la\sim 100~{\rm nm}$, $\eta\sim1$ (appropriate for NiO and NiO$|$FeNi interface) and $L\sim10~{\rm nm}$, we obtain $\beta_{\rm th}\approx0.8$, which is quantitatively consistent with the experiment~\cite{moriyamaCM15}. 

Our results can be used to make a connection with the reciprocal experiments~\cite{hahnEPL14,wangPRL14}, in which spin transfer through the AF is quantified by measuring the inverse spin Hall voltage $V_{\rm ISHE}$ generated across N by a dynamic F (see Fig.~\ref{setup}). From spin Hall phenomenology and Onsager reciprocity, the electromotive force generated in N is given by $\boldsymbol{\ve}=\vartheta(\bn\times\dot\bn)\times\ex|_{x=0}$~\cite{tserkovnyakPRB14}. Utilizing the adiabatic result $\bn^{(0)}(x,t)$ [i.e., Eq.~\eqref{n0}] with $\bj=0$, the (time-averaged) motive force becomes $\langle\boldsymbol{\ve}\rangle=-(\vartheta^*\theta^2b_0/\hbar)\ey$, where $\theta\approx (s_x^2+s_y^2)^{1/2}$ is the cone angle, $\vartheta^*=\vartheta/\{\sV_F[\cosh(L/\la)+\sinh(L/\la)/\eta]^2\}$, and $\sV_F$ is the volume of F. This leads to an inverse spin Hall voltage
\beq
\label{vishe}
V_{\rm ISHE}=-\frac{\vartheta\theta^2b_0\ell}{\hbar \sV_F\round{\cosh\frac{L}{\la}+\frac{1}{\eta}\sinh\frac{L}{\la}}^2}\, ,
\eeq 
where $\ell$ is the length of N in the $y$ direction. We can arrive at the same result by treating the central AF as an effective junction between the N and F subsystems (see Fig.~\ref{setup}). Namely, from Eq.~\eqref{dali}, the macroscopic coupling between current $\bj$ and the macrospin dynamics in F is given through the torque $\boldsymbol{\tau}=\vartheta^*\theta^2j_y\ez+({\rm term }\propto j_z)$ acting on the latter, where $\vartheta^*$ is the overall torque coefficient for the effective junction. By Onsager reciprocity, this torque gives rise to the inverse spin Hall voltage Eq.~\eqref{vishe}. Within the current theory, Eq.~\eqref{vishe} indicates that the decay length for $V_{\rm ISHE}$ as the AF thickness increases is set by $\la$.

We note in conclusion that the current work considers spin transfer purely mediated by the coherent N\'eel dynamics, corresponding to the so-called superfluid contribution to spin transport. As the relevant experiments are performed at room temperature~\cite{hahnEPL14,wangPRL14,moriyamaCM15} reconsidering AF spin transport by accounting for the incoherent thermal magnons and studying their effect on the FMR linewidth would be valuable, and will contribute to the general understanding of the ``two-fluid" (condensate and thermal cloud with mutual interactions between them) nature of spin transport via collective excitations in an AF.

The authors would like to thank P. Chris Hammel and Fengyuan Yang for useful discussions. This work was supported by FAME (an SRC STARnet center sponsored by MARCO and DARPA) (S.T. and Y.T.) and in part by Grants-in-Aid for Scientific Research (S) and Grant-in-Aid for Young Scientists (B) from the Japan Society for the Promotion of Science (T.M. and T.O.).


%


\begin{thebibliography}{31}%
\makeatletter
\providecommand \@ifxundefined [1]{%
 \@ifx{#1\undefined}
}%
\providecommand \@ifnum [1]{%
 \ifnum #1\expandafter \@firstoftwo
 \else \expandafter \@secondoftwo
 \fi
}%
\providecommand \@ifx [1]{%
 \ifx #1\expandafter \@firstoftwo
 \else \expandafter \@secondoftwo
 \fi
}%
\providecommand \natexlab [1]{#1}%
\providecommand \enquote  [1]{``#1''}%
\providecommand \bibnamefont  [1]{#1}%
\providecommand \bibfnamefont [1]{#1}%
\providecommand \citenamefont [1]{#1}%
\providecommand \href@noop [0]{\@secondoftwo}%
\providecommand \href [0]{\begingroup \@sanitize@url \@href}%
\providecommand \@href[1]{\@@startlink{#1}\@@href}%
\providecommand \@@href[1]{\endgroup#1\@@endlink}%
\providecommand \@sanitize@url [0]{\catcode `\\12\catcode `\$12\catcode
  `\&12\catcode `\#12\catcode `\^12\catcode `\_12\catcode `\%12\relax}%
\providecommand \@@startlink[1]{}%
\providecommand \@@endlink[0]{}%
\providecommand \url  [0]{\begingroup\@sanitize@url \@url }%
\providecommand \@url [1]{\endgroup\@href {#1}{\urlprefix }}%
\providecommand \urlprefix  [0]{URL }%
\providecommand \Eprint [0]{\href }%
\providecommand \doibase [0]{http://dx.doi.org/}%
\providecommand \selectlanguage [0]{\@gobble}%
\providecommand \bibinfo  [0]{\@secondoftwo}%
\providecommand \bibfield  [0]{\@secondoftwo}%
\providecommand \translation [1]{[#1]}%
\providecommand \BibitemOpen [0]{}%
\providecommand \bibitemStop [0]{}%
\providecommand \bibitemNoStop [0]{.\EOS\space}%
\providecommand \EOS [0]{\spacefactor3000\relax}%
\providecommand \BibitemShut  [1]{\csname bibitem#1\endcsname}%
\let\auto@bib@innerbib\@empty
\bibitem [{\citenamefont {MacDonald}\ and\ \citenamefont
  {Tsoi}(2011)}]{macdonalPTRSA11}%
  \BibitemOpen
  \bibfield  {author} {\bibinfo {author} {\bibfnamefont {A.~H.}\ \bibnamefont
  {MacDonald}}\ and\ \bibinfo {author} {\bibfnamefont {M.}~\bibnamefont
  {Tsoi}},\ }\href {\doibase 10.1098/rsta.2011.0014} {\bibfield  {journal}
  {\bibinfo  {journal} {Philosophical Transactions of the Royal Society A}\
  }\textbf {\bibinfo {volume} {369}},\ \bibinfo {pages} {3098} (\bibinfo {year}
  {2011})}\BibitemShut {NoStop}%
\bibitem [{\citenamefont {Duine}(2011)}]{duineNATM11}%
  \BibitemOpen
  \bibfield  {author} {\bibinfo {author} {\bibfnamefont {R.}~\bibnamefont
  {Duine}},\ }\href {http://dx.doi.org/10.1038/nmat3015} {\bibfield  {journal}
  {\bibinfo  {journal} {Nat Mater}\ }\textbf {\bibinfo {volume} {10}},\
  \bibinfo {pages} {344} (\bibinfo {year} {2011})}\BibitemShut {NoStop}%
\bibitem [{\citenamefont {Brataas}\ \emph {et~al.}(2012)\citenamefont
  {Brataas}, \citenamefont {Kent},\ and\ \citenamefont {Ohno}}]{brataasNATM12}%
  \BibitemOpen
  \bibfield  {author} {\bibinfo {author} {\bibfnamefont {A.}~\bibnamefont
  {Brataas}}, \bibinfo {author} {\bibfnamefont {A.~D.}\ \bibnamefont {Kent}}, \
  and\ \bibinfo {author} {\bibfnamefont {H.}~\bibnamefont {Ohno}},\ }\href@noop
  {} {\bibfield  {journal} {\bibinfo  {journal} {Nat Mater}\ }\textbf {\bibinfo
  {volume} {11}},\ \bibinfo {pages} {372} (\bibinfo {year} {2012})}\BibitemShut
  {NoStop}%
\bibitem [{\citenamefont {N{\'u}{\~n}ez}\ \emph {et~al.}(2006)\citenamefont
  {N{\'u}{\~n}ez}, \citenamefont {Duine}, \citenamefont {Haney},\ and\
  \citenamefont {MacDonald}}]{nunezPRB06}%
  \BibitemOpen
  \bibfield  {author} {\bibinfo {author} {\bibfnamefont {A.~S.}\ \bibnamefont
  {N{\'u}{\~n}ez}}, \bibinfo {author} {\bibfnamefont {R.~A.}\ \bibnamefont
  {Duine}}, \bibinfo {author} {\bibfnamefont {P.}~\bibnamefont {Haney}}, \ and\
  \bibinfo {author} {\bibfnamefont {A.~H.}\ \bibnamefont {MacDonald}},\
  }\href@noop {} {\bibfield  {journal} {\bibinfo  {journal} {Phys. Rev. B}\
  }\textbf {\bibinfo {volume} {73}},\ \bibinfo {eid} {214426} (\bibinfo {year}
  {2006})}\BibitemShut {NoStop}%
\bibitem [{\citenamefont {Haney}\ \emph {et~al.}(2007)\citenamefont {Haney},
  \citenamefont {Waldron}, \citenamefont {Duine}, \citenamefont {N\'u\~nez},
  \citenamefont {Guo},\ and\ \citenamefont {MacDonald}}]{haneyPRB07}%
  \BibitemOpen
  \bibfield  {author} {\bibinfo {author} {\bibfnamefont {P.~M.}\ \bibnamefont
  {Haney}}, \bibinfo {author} {\bibfnamefont {D.}~\bibnamefont {Waldron}},
  \bibinfo {author} {\bibfnamefont {R.~A.}\ \bibnamefont {Duine}}, \bibinfo
  {author} {\bibfnamefont {A.~S.}\ \bibnamefont {N\'u\~nez}}, \bibinfo {author}
  {\bibfnamefont {H.}~\bibnamefont {Guo}}, \ and\ \bibinfo {author}
  {\bibfnamefont {A.~H.}\ \bibnamefont {MacDonald}},\ }\href {\doibase
  10.1103/PhysRevB.75.174428} {\bibfield  {journal} {\bibinfo  {journal} {Phys.
  Rev. B}\ }\textbf {\bibinfo {volume} {75}},\ \bibinfo {pages} {174428}
  (\bibinfo {year} {2007})}\BibitemShut {NoStop}%
\bibitem [{\citenamefont {Urazhdin}\ and\ \citenamefont
  {Anthony}(2007)}]{urazhdinPRL07}%
  \BibitemOpen
  \bibfield  {author} {\bibinfo {author} {\bibfnamefont {S.}~\bibnamefont
  {Urazhdin}}\ and\ \bibinfo {author} {\bibfnamefont {N.}~\bibnamefont
  {Anthony}},\ }\href@noop {} {\bibfield  {journal} {\bibinfo  {journal} {Phys.
  Rev. Lett.}\ }\textbf {\bibinfo {volume} {99}},\ \bibinfo {eid} {046602}
  (\bibinfo {year} {2007})}\BibitemShut {NoStop}%
\bibitem [{\citenamefont {Wei}\ \emph {et~al.}(2007)\citenamefont {Wei},
  \citenamefont {Sharma}, \citenamefont {Nunez}, \citenamefont {Haney},
  \citenamefont {Duine}, \citenamefont {Bass}, \citenamefont {MacDonald},\ and\
  \citenamefont {Tsoi}}]{weiPRL07}%
  \BibitemOpen
  \bibfield  {author} {\bibinfo {author} {\bibfnamefont {Z.}~\bibnamefont
  {Wei}}, \bibinfo {author} {\bibfnamefont {A.}~\bibnamefont {Sharma}},
  \bibinfo {author} {\bibfnamefont {A.~S.}\ \bibnamefont {Nunez}}, \bibinfo
  {author} {\bibfnamefont {P.~M.}\ \bibnamefont {Haney}}, \bibinfo {author}
  {\bibfnamefont {R.~A.}\ \bibnamefont {Duine}}, \bibinfo {author}
  {\bibfnamefont {J.}~\bibnamefont {Bass}}, \bibinfo {author} {\bibfnamefont
  {A.~H.}\ \bibnamefont {MacDonald}}, \ and\ \bibinfo {author} {\bibfnamefont
  {M.}~\bibnamefont {Tsoi}},\ }\href {\doibase 10.1103/PhysRevLett.98.116603}
  {\bibfield  {journal} {\bibinfo  {journal} {Phys. Rev. Lett.}\ }\textbf
  {\bibinfo {volume} {98}},\ \bibinfo {pages} {116603} (\bibinfo {year}
  {2007})}\BibitemShut {NoStop}%
\bibitem [{\citenamefont {Haney}\ and\ \citenamefont
  {MacDonald}(2008)}]{haneyPRL08}%
  \BibitemOpen
  \bibfield  {author} {\bibinfo {author} {\bibfnamefont {P.~M.}\ \bibnamefont
  {Haney}}\ and\ \bibinfo {author} {\bibfnamefont {A.~H.}\ \bibnamefont
  {MacDonald}},\ }\href {\doibase 10.1103/PhysRevLett.100.196801} {\bibfield
  {journal} {\bibinfo  {journal} {Phys. Rev. Lett.}\ }\textbf {\bibinfo
  {volume} {100}},\ \bibinfo {pages} {196801} (\bibinfo {year}
  {2008})}\BibitemShut {NoStop}%
\bibitem [{\citenamefont {Xu}\ \emph {et~al.}(2008)\citenamefont {Xu},
  \citenamefont {Wang},\ and\ \citenamefont {Xia}}]{xuPRL08}%
  \BibitemOpen
  \bibfield  {author} {\bibinfo {author} {\bibfnamefont {Y.}~\bibnamefont
  {Xu}}, \bibinfo {author} {\bibfnamefont {S.}~\bibnamefont {Wang}}, \ and\
  \bibinfo {author} {\bibfnamefont {K.}~\bibnamefont {Xia}},\ }\href@noop {}
  {\bibfield  {journal} {\bibinfo  {journal} {Phys. Rev. Lett.}\ }\textbf
  {\bibinfo {volume} {100}},\ \bibinfo {eid} {226602} (\bibinfo {year}
  {2008})}\BibitemShut {NoStop}%
\bibitem [{\citenamefont {Gomonay}\ and\ \citenamefont
  {Loktev}(2010)}]{gomonayPRB10}%
  \BibitemOpen
  \bibfield  {author} {\bibinfo {author} {\bibfnamefont {H.~V.}\ \bibnamefont
  {Gomonay}}\ and\ \bibinfo {author} {\bibfnamefont {V.~M.}\ \bibnamefont
  {Loktev}},\ }\href {\doibase 10.1103/PhysRevB.81.144427} {\bibfield
  {journal} {\bibinfo  {journal} {Phys. Rev. B}\ }\textbf {\bibinfo {volume}
  {81}},\ \bibinfo {pages} {144427} (\bibinfo {year} {2010})}\BibitemShut
  {NoStop}%
\bibitem [{\citenamefont {Gomonay}\ \emph {et~al.}(2012)\citenamefont
  {Gomonay}, \citenamefont {Kunitsyn},\ and\ \citenamefont
  {Loktev}}]{gomonayPRB12}%
  \BibitemOpen
  \bibfield  {author} {\bibinfo {author} {\bibfnamefont {H.~V.}\ \bibnamefont
  {Gomonay}}, \bibinfo {author} {\bibfnamefont {R.~V.}\ \bibnamefont
  {Kunitsyn}}, \ and\ \bibinfo {author} {\bibfnamefont {V.~M.}\ \bibnamefont
  {Loktev}},\ }\href {\doibase 10.1103/PhysRevB.85.134446} {\bibfield
  {journal} {\bibinfo  {journal} {Phys. Rev. B}\ }\textbf {\bibinfo {volume}
  {85}},\ \bibinfo {pages} {134446} (\bibinfo {year} {2012})}\BibitemShut
  {NoStop}%
\bibitem [{\citenamefont {Shick}\ \emph {et~al.}(2010)\citenamefont {Shick},
  \citenamefont {Khmelevskyi}, \citenamefont {Mryasov}, \citenamefont
  {Wunderlich},\ and\ \citenamefont {Jungwirth}}]{shickPRB10}%
  \BibitemOpen
  \bibfield  {author} {\bibinfo {author} {\bibfnamefont {A.~B.}\ \bibnamefont
  {Shick}}, \bibinfo {author} {\bibfnamefont {S.}~\bibnamefont {Khmelevskyi}},
  \bibinfo {author} {\bibfnamefont {O.~N.}\ \bibnamefont {Mryasov}}, \bibinfo
  {author} {\bibfnamefont {J.}~\bibnamefont {Wunderlich}}, \ and\ \bibinfo
  {author} {\bibfnamefont {T.}~\bibnamefont {Jungwirth}},\ }\href {\doibase
  10.1103/PhysRevB.81.212409} {\bibfield  {journal} {\bibinfo  {journal} {Phys.
  Rev. B}\ }\textbf {\bibinfo {volume} {81}},\ \bibinfo {pages} {212409}
  (\bibinfo {year} {2010})}\BibitemShut {NoStop}%
\bibitem [{\citenamefont {Park}\ \emph {et~al.}(2011)\citenamefont {Park},
  \citenamefont {Wunderlich}, \citenamefont {Mart\'i}, \citenamefont {Hol\'y},
  \citenamefont {Kurosaki}, \citenamefont {Yamada}, \citenamefont {Yamamoto},
  \citenamefont {Nishide}, \citenamefont {Hayakawa}, \citenamefont {Takahashi},
  \citenamefont {Shick},\ and\ \citenamefont {Jungwirth}}]{parkNATM11}%
  \BibitemOpen
  \bibfield  {author} {\bibinfo {author} {\bibfnamefont {B.~G.}\ \bibnamefont
  {Park}}, \bibinfo {author} {\bibfnamefont {J.}~\bibnamefont {Wunderlich}},
  \bibinfo {author} {\bibfnamefont {X.}~\bibnamefont {Mart\'i}}, \bibinfo
  {author} {\bibfnamefont {V.}~\bibnamefont {Hol\'y}}, \bibinfo {author}
  {\bibfnamefont {Y.}~\bibnamefont {Kurosaki}}, \bibinfo {author}
  {\bibfnamefont {M.}~\bibnamefont {Yamada}}, \bibinfo {author} {\bibfnamefont
  {H.}~\bibnamefont {Yamamoto}}, \bibinfo {author} {\bibfnamefont
  {A.}~\bibnamefont {Nishide}}, \bibinfo {author} {\bibfnamefont
  {J.}~\bibnamefont {Hayakawa}}, \bibinfo {author} {\bibfnamefont
  {H.}~\bibnamefont {Takahashi}}, \bibinfo {author} {\bibfnamefont {A.~B.}\
  \bibnamefont {Shick}}, \ and\ \bibinfo {author} {\bibfnamefont
  {T.}~\bibnamefont {Jungwirth}},\ }\href@noop {} {\bibfield  {journal}
  {\bibinfo  {journal} {Nat Mater}\ }\textbf {\bibinfo {volume} {10}},\
  \bibinfo {pages} {347} (\bibinfo {year} {2011})}\BibitemShut {NoStop}%
\bibitem [{\citenamefont {K{\"o}nig}\ \emph {et~al.}(2001)\citenamefont
  {K{\"o}nig}, \citenamefont {B{\o}nsager},\ and\ \citenamefont
  {MacDonald}}]{konigPRL01}%
  \BibitemOpen
  \bibfield  {author} {\bibinfo {author} {\bibfnamefont {J.}~\bibnamefont
  {K{\"o}nig}}, \bibinfo {author} {\bibfnamefont {M.~C.}\ \bibnamefont
  {B{\o}nsager}}, \ and\ \bibinfo {author} {\bibfnamefont {A.~H.}\ \bibnamefont
  {MacDonald}},\ }\href@noop {} {\bibfield  {journal} {\bibinfo  {journal}
  {Phys. Rev. Lett.}\ }\textbf {\bibinfo {volume} {87}},\ \bibinfo {pages}
  {187202} (\bibinfo {year} {2001})}\BibitemShut {NoStop}%
\bibitem [{\citenamefont {Sonin}(2010)}]{soninAP10}%
  \BibitemOpen
  \bibfield  {author} {\bibinfo {author} {\bibfnamefont {E.~B.}\ \bibnamefont
  {Sonin}},\ }\href@noop {} {\bibfield  {journal} {\bibinfo  {journal} {Adv.
  Phys.}\ }\textbf {\bibinfo {volume} {59}},\ \bibinfo {pages} {181} (\bibinfo
  {year} {2010})}\BibitemShut {NoStop}%
\bibitem [{\citenamefont {Takei}\ \emph {et~al.}(2014)\citenamefont {Takei},
  \citenamefont {Halperin}, \citenamefont {Yacoby},\ and\ \citenamefont
  {Tserkovnyak}}]{takeiPRB14}%
  \BibitemOpen
  \bibfield  {author} {\bibinfo {author} {\bibfnamefont {S.}~\bibnamefont
  {Takei}}, \bibinfo {author} {\bibfnamefont {B.~I.}\ \bibnamefont {Halperin}},
  \bibinfo {author} {\bibfnamefont {A.}~\bibnamefont {Yacoby}}, \ and\ \bibinfo
  {author} {\bibfnamefont {Y.}~\bibnamefont {Tserkovnyak}},\ }\href {\doibase
  10.1103/PhysRevB.90.094408} {\bibfield  {journal} {\bibinfo  {journal} {Phys.
  Rev. B}\ }\textbf {\bibinfo {volume} {90}},\ \bibinfo {pages} {094408}
  (\bibinfo {year} {2014})}\BibitemShut {NoStop}%
\bibitem [{\citenamefont {Wieser}\ \emph {et~al.}(2011)\citenamefont {Wieser},
  \citenamefont {Vedmedenko},\ and\ \citenamefont
  {Wiesendanger}}]{wieserPRL11}%
  \BibitemOpen
  \bibfield  {author} {\bibinfo {author} {\bibfnamefont {R.}~\bibnamefont
  {Wieser}}, \bibinfo {author} {\bibfnamefont {E.~Y.}\ \bibnamefont
  {Vedmedenko}}, \ and\ \bibinfo {author} {\bibfnamefont {R.}~\bibnamefont
  {Wiesendanger}},\ }\href {\doibase 10.1103/PhysRevLett.106.067204} {\bibfield
   {journal} {\bibinfo  {journal} {Phys. Rev. Lett.}\ }\textbf {\bibinfo
  {volume} {106}},\ \bibinfo {pages} {067204} (\bibinfo {year}
  {2011})}\BibitemShut {NoStop}%
\bibitem [{\citenamefont {Kim}\ \emph {et~al.}(2014)\citenamefont {Kim},
  \citenamefont {Tserkovnyak},\ and\ \citenamefont {Tchernyshyov}}]{kimPRB14}%
  \BibitemOpen
  \bibfield  {author} {\bibinfo {author} {\bibfnamefont {S.~K.}\ \bibnamefont
  {Kim}}, \bibinfo {author} {\bibfnamefont {Y.}~\bibnamefont {Tserkovnyak}}, \
  and\ \bibinfo {author} {\bibfnamefont {O.}~\bibnamefont {Tchernyshyov}},\
  }\href {\doibase 10.1103/PhysRevB.90.104406} {\bibfield  {journal} {\bibinfo
  {journal} {Phys. Rev. B}\ }\textbf {\bibinfo {volume} {90}},\ \bibinfo
  {pages} {104406} (\bibinfo {year} {2014})}\BibitemShut {NoStop}%
\bibitem [{\citenamefont {Hals}\ \emph {et~al.}(2011)\citenamefont {Hals},
  \citenamefont {Tserkovnyak},\ and\ \citenamefont {Brataas}}]{halsPRL11}%
  \BibitemOpen
  \bibfield  {author} {\bibinfo {author} {\bibfnamefont {K.~M.~D.}\
  \bibnamefont {Hals}}, \bibinfo {author} {\bibfnamefont {Y.}~\bibnamefont
  {Tserkovnyak}}, \ and\ \bibinfo {author} {\bibfnamefont {A.}~\bibnamefont
  {Brataas}},\ }\href {\doibase 10.1103/PhysRevLett.106.107206} {\bibfield
  {journal} {\bibinfo  {journal} {Phys. Rev. Lett.}\ }\textbf {\bibinfo
  {volume} {106}},\ \bibinfo {pages} {107206} (\bibinfo {year}
  {2011})}\BibitemShut {NoStop}%
\bibitem [{\citenamefont {Swaving}\ and\ \citenamefont
  {Duine}(2011)}]{swavingPRB11}%
  \BibitemOpen
  \bibfield  {author} {\bibinfo {author} {\bibfnamefont {A.~C.}\ \bibnamefont
  {Swaving}}\ and\ \bibinfo {author} {\bibfnamefont {R.~A.}\ \bibnamefont
  {Duine}},\ }\href {\doibase 10.1103/PhysRevB.83.054428} {\bibfield  {journal}
  {\bibinfo  {journal} {Phys. Rev. B}\ }\textbf {\bibinfo {volume} {83}},\
  \bibinfo {pages} {054428} (\bibinfo {year} {2011})}\BibitemShut {NoStop}%
\bibitem [{\citenamefont {Cheng}\ and\ \citenamefont {Niu}(2012)}]{chengPRB12}%
  \BibitemOpen
  \bibfield  {author} {\bibinfo {author} {\bibfnamefont {R.}~\bibnamefont
  {Cheng}}\ and\ \bibinfo {author} {\bibfnamefont {Q.}~\bibnamefont {Niu}},\
  }\href {\doibase 10.1103/PhysRevB.86.245118} {\bibfield  {journal} {\bibinfo
  {journal} {Phys. Rev. B}\ }\textbf {\bibinfo {volume} {86}},\ \bibinfo
  {pages} {245118} (\bibinfo {year} {2012})}\BibitemShut {NoStop}%
\bibitem [{\citenamefont {Tveten}\ \emph {et~al.}(2013)\citenamefont {Tveten},
  \citenamefont {Qaiumzadeh}, \citenamefont {Tretiakov},\ and\ \citenamefont
  {Brataas}}]{tvetenPRL13}%
  \BibitemOpen
  \bibfield  {author} {\bibinfo {author} {\bibfnamefont {E.~G.}\ \bibnamefont
  {Tveten}}, \bibinfo {author} {\bibfnamefont {A.}~\bibnamefont {Qaiumzadeh}},
  \bibinfo {author} {\bibfnamefont {O.~A.}\ \bibnamefont {Tretiakov}}, \ and\
  \bibinfo {author} {\bibfnamefont {A.}~\bibnamefont {Brataas}},\ }\href
  {\doibase 10.1103/PhysRevLett.110.127208} {\bibfield  {journal} {\bibinfo
  {journal} {Phys. Rev. Lett.}\ }\textbf {\bibinfo {volume} {110}},\ \bibinfo
  {pages} {127208} (\bibinfo {year} {2013})}\BibitemShut {NoStop}%
\bibitem [{\citenamefont {Hahn}\ \emph {et~al.}(2014)\citenamefont {Hahn},
  \citenamefont {de~Loubens}, \citenamefont {Naletov}, \citenamefont {Youssef},
  \citenamefont {Klein},\ and\ \citenamefont {Viret}}]{hahnEPL14}%
  \BibitemOpen
  \bibfield  {author} {\bibinfo {author} {\bibfnamefont {C.}~\bibnamefont
  {Hahn}}, \bibinfo {author} {\bibfnamefont {G.}~\bibnamefont {de~Loubens}},
  \bibinfo {author} {\bibfnamefont {V.~V.}\ \bibnamefont {Naletov}}, \bibinfo
  {author} {\bibfnamefont {J.~B.}\ \bibnamefont {Youssef}}, \bibinfo {author}
  {\bibfnamefont {O.}~\bibnamefont {Klein}}, \ and\ \bibinfo {author}
  {\bibfnamefont {M.}~\bibnamefont {Viret}},\ }\href
  {http://stacks.iop.org/0295-5075/108/i=5/a=57005} {\bibfield  {journal}
  {\bibinfo  {journal} {EPL (Europhysics Letters)}\ }\textbf {\bibinfo {volume}
  {108}},\ \bibinfo {pages} {57005} (\bibinfo {year} {2014})}\BibitemShut
  {NoStop}%
\bibitem [{\citenamefont {Wang}\ \emph {et~al.}(2014)\citenamefont {Wang},
  \citenamefont {Du}, \citenamefont {Hammel},\ and\ \citenamefont
  {Yang}}]{wangPRL14}%
  \BibitemOpen
  \bibfield  {author} {\bibinfo {author} {\bibfnamefont {H.}~\bibnamefont
  {Wang}}, \bibinfo {author} {\bibfnamefont {C.}~\bibnamefont {Du}}, \bibinfo
  {author} {\bibfnamefont {P.~C.}\ \bibnamefont {Hammel}}, \ and\ \bibinfo
  {author} {\bibfnamefont {F.}~\bibnamefont {Yang}},\ }\href {\doibase
  10.1103/PhysRevLett.113.097202} {\bibfield  {journal} {\bibinfo  {journal}
  {Phys. Rev. Lett.}\ }\textbf {\bibinfo {volume} {113}},\ \bibinfo {pages}
  {097202} (\bibinfo {year} {2014})}\BibitemShut {NoStop}%
\bibitem [{\citenamefont {{Moriyama}}\ \emph {et~al.}(2014)\citenamefont
  {{Moriyama}}, \citenamefont {{Nagata}}, \citenamefont {{Tanaka}},
  \citenamefont {{Kim}}, \citenamefont {{Almasi}}, \citenamefont {{Wang}},\
  and\ \citenamefont {{Ono}}}]{moriyamaCM14}%
  \BibitemOpen
  \bibfield  {author} {\bibinfo {author} {\bibfnamefont {T.}~\bibnamefont
  {{Moriyama}}}, \bibinfo {author} {\bibfnamefont {M.}~\bibnamefont
  {{Nagata}}}, \bibinfo {author} {\bibfnamefont {K.}~\bibnamefont {{Tanaka}}},
  \bibinfo {author} {\bibfnamefont {K.}~\bibnamefont {{Kim}}}, \bibinfo
  {author} {\bibfnamefont {H.}~\bibnamefont {{Almasi}}}, \bibinfo {author}
  {\bibfnamefont {W.}~\bibnamefont {{Wang}}}, \ and\ \bibinfo {author}
  {\bibfnamefont {T.}~\bibnamefont {{Ono}}},\ }\href@noop {} {\bibfield
  {journal} {\bibinfo  {journal} {ArXiv e-prints}\ } (\bibinfo {year}
  {2014})},\ \Eprint {http://arxiv.org/abs/1411.4100} {arXiv:1411.4100
  [cond-mat.mtrl-sci]} \BibitemShut {NoStop}%
\bibitem [{\citenamefont {{Moriyama}}\ \emph {et~al.}()\citenamefont
  {{Moriyama}}, \citenamefont {{Takei}}, \citenamefont {{Nagata}},
  \citenamefont {{Yoshimura}}, \citenamefont {{Matsuzaki}}, \citenamefont
  {{Terashima}}, \citenamefont {{Tserkovnyak}},\ and\ \citenamefont
  {{Ono}}}]{moriyamaCM15}%
  \BibitemOpen
  \bibfield  {author} {\bibinfo {author} {\bibfnamefont {T.}~\bibnamefont
  {{Moriyama}}}, \bibinfo {author} {\bibfnamefont {S.}~\bibnamefont {{Takei}}},
  \bibinfo {author} {\bibfnamefont {M.}~\bibnamefont {{Nagata}}}, \bibinfo
  {author} {\bibfnamefont {Y.}~\bibnamefont {{Yoshimura}}}, \bibinfo {author}
  {\bibfnamefont {N.}~\bibnamefont {{Matsuzaki}}}, \bibinfo {author}
  {\bibfnamefont {T.}~\bibnamefont {{Terashima}}}, \bibinfo {author}
  {\bibfnamefont {Y.}~\bibnamefont {{Tserkovnyak}}}, \ and\ \bibinfo {author}
  {\bibfnamefont {T.}~\bibnamefont {{Ono}}},\ }\href@noop {} {\bibinfo
  {journal} {in preparation}\ }\BibitemShut {NoStop}%
\bibitem [{\citenamefont {Bar'yakhtar}\ and\ \citenamefont
  {Ivanov}(1979)}]{baryakhtarSJLTP79}%
  \BibitemOpen
\bibfield  {journal} {  }\bibfield  {author} {\bibinfo {author} {\bibfnamefont
  {V.~G.}\ \bibnamefont {Bar'yakhtar}}\ and\ \bibinfo {author} {\bibfnamefont
  {B.~A.}\ \bibnamefont {Ivanov}},\ }\href@noop {} {\bibfield  {journal}
  {\bibinfo  {journal} {Sov. J. Low. Temp. Phys.}\ }\textbf {\bibinfo {volume}
  {5}},\ \bibinfo {pages} {361} (\bibinfo {year} {1979})}\BibitemShut {NoStop}%
\bibitem [{\citenamefont {Haldane}(1983)}]{haldanePRL83}%
  \BibitemOpen
  \bibfield  {author} {\bibinfo {author} {\bibfnamefont {F.~D.~M.}\
  \bibnamefont {Haldane}},\ }\href@noop {} {\bibfield  {journal} {\bibinfo
  {journal} {Phys. Rev. Lett.}\ }\textbf {\bibinfo {volume} {51}},\ \bibinfo
  {pages} {605} (\bibinfo {year} {1983})}\BibitemShut {NoStop}%
\bibitem [{\citenamefont {Auerbach}(1994)}]{auerbachBOOK94}%
  \BibitemOpen
  \bibfield  {author} {\bibinfo {author} {\bibfnamefont {A.}~\bibnamefont
  {Auerbach}},\ }\href@noop {} {\emph {\bibinfo {title} {Interacting Electrons
  and Quantum Magnetism}}}\ (\bibinfo  {publisher} {Springer-Verlag},\ \bibinfo
  {address} {New York},\ \bibinfo {year} {1994})\BibitemShut {NoStop}%
\bibitem [{Note1()}]{Note1}%
  \BibitemOpen
  \bibinfo {note} {The saturated spin density is defined by $s=\hbar
  S_0\protect \mathscr {A}/\protect \mathscr {V}$, where $S_0$ is the magnitude
  of each AF spin, $\protect \mathscr {A}$ is the cross-sectional area (along
  the $yz$ plane) of the AF, and $\protect \mathscr {V}$ is the volume occupied
  by each spin.}\BibitemShut {Stop}%
\bibitem [{\citenamefont {Tserkovnyak}\ and\ \citenamefont
  {Bender}(2014)}]{tserkovnyakPRB14}%
  \BibitemOpen
  \bibfield  {author} {\bibinfo {author} {\bibfnamefont {Y.}~\bibnamefont
  {Tserkovnyak}}\ and\ \bibinfo {author} {\bibfnamefont {S.~A.}\ \bibnamefont
  {Bender}},\ }\href {\doibase 10.1103/PhysRevB.90.014428} {\bibfield
  {journal} {\bibinfo  {journal} {Phys. Rev. B}\ }\textbf {\bibinfo {volume}
  {90}},\ \bibinfo {pages} {014428} (\bibinfo {year} {2014})}\BibitemShut
  {NoStop}%
\end{thebibliography}
\end{document}